\documentclass[aps,prl,twocolumn,showpacs,superscriptaddress,groupedaddress]{revtex4-1}

\usepackage[utf8x]{inputenc}
\usepackage{graphicx}
\usepackage{dcolumn}
\usepackage{bm}
\usepackage{amsmath}
\usepackage{amssymb}
\usepackage{subfigure}
\usepackage{subfloat}

\begin{document}

\title{Compact FPGA-based pulse-sequencer and radio-frequency generator for experiments with trapped atoms}

\author{Thaned Pruttivarasin}
\email{thaned.pruttivarasin@riken.jp}
\affiliation{Quantum Metrology Laboratory, RIKEN, Wako-shi, Saitama 351-0198, Japan}

\author{Hidetoshi Katori}
\affiliation{Quantum Metrology Laboratory, RIKEN, Wako-shi, Saitama 351-0198, Japan}
\affiliation{Innovative Space-Time Project, ERATO, JST, Bunkyo-ku, Tokyo 113-8656, Japan}
\affiliation{Department of Applied Physics, Graduate School of Engineering,
The University of Tokyo, Bunkyo-ku, Tokyo 113-8656, Japan}

\date{\today}

\begin{abstract}

We present a compact FPGA-based pulse sequencer and radio-frequency (RF) generator suitable for experiments with cold trapped ions and atoms. The unit is capable of outputting a pulse sequence with at least 32 TTL channels with a timing resolution of 40 ns and contains a built-in 100 MHz frequency counter for counting electrical pulses from a photo-multiplier tube (PMT). There are 16 independent direct-digital-synthesizers (DDS) RF sources with fast (rise-time of $\sim$60 ns) amplitude switching and sub-mHz frequency tuning from 0 to 800 MHz.

\end{abstract}

\maketitle

\section{Introduction}

Experiments with cold trapped atoms rely on precise control of laser light amplitude and frequency. Depending on the details of the experiments, the duration of the laser pulse can range from a few seconds to a few micro-seconds. To meet these timing requirements, switching of laser light is usually achieved by a combination of acousto-optical modulators (AOMs) for fast switching and precise frequency tuning, and mechanical shutters to eliminate any leakage of the laser light. Since AOMs require radio-frequency (RF) signals (typically with frequency from 0 to 500 MHz) to operate, RF generators are indispensable in every modern atomic physics laboratory.

While there are many commercially available RF generators in the market, they are not optimized to control multiple AOMs in atomic physics experiments. For example, most RF generators requires a few milli-seconds to reprogram the frequency and/or the amplitude. Some devices have a frequency-shift-key (FSK) functionality, but that only allows users to quickly alternate between two fixed frequencies. To overcome this limitation, multiple RF generators combined with RF switches can be used. As the number of laser sources increases with the complexity of the experiment, the space these devices take up in the laboratory can be significant. Moreover, incautious wiring among these devices can lead to unwanted electromagnetic interferences between RF sources and other electronics. Hence, a compact multi-channel RF source with robust frequency and amplitude control is desirable since it simplifies the experimental setup.

In this paper, we report on our development of a compact FPGA-based multi-channel RF-generator and pulse sequencer \cite{history}. The system contains a multi-channel TTL digital pulse sequencer with a timing resolution of 40 ns and 16 channels of direct-digital synthesized (DDS) RF generator with a frequency tuning resolution of better than 1 mHz, which is especially beneficial in operating optical atomic clocks. Each DDS channel is capable of amplitude switching with a rise-time of 60 ns and consecutively switching RF frequency within 1.0 $\mu$s. Additionally, users can independently program amplitude and frequency ramps of each DDS channel independently, making the unit suitable for a wide range of experiments. The unit also has a built-in frequency counter for counting electrical pulses from photo-multiplier tube (PMT), which is widely used to detect fluorescence from trapped atoms or ions. The counter is capable of time-tagging the arrival of the photons at the PMT referenced to the timing of the pulse sequence with a resolution of 10 ns. Only a single USB cable is required between the pulse sequencer and a control computer. The system (not including power supplies and 2 GHz reference clocks) takes up a volume of $35\times23\times13$ cm$^3$ which fits in a 3U 19-inch standard rack.

\section{Design concept}

\subsection{Overview}

\begin{figure*}
\includegraphics[width=0.95\textwidth]{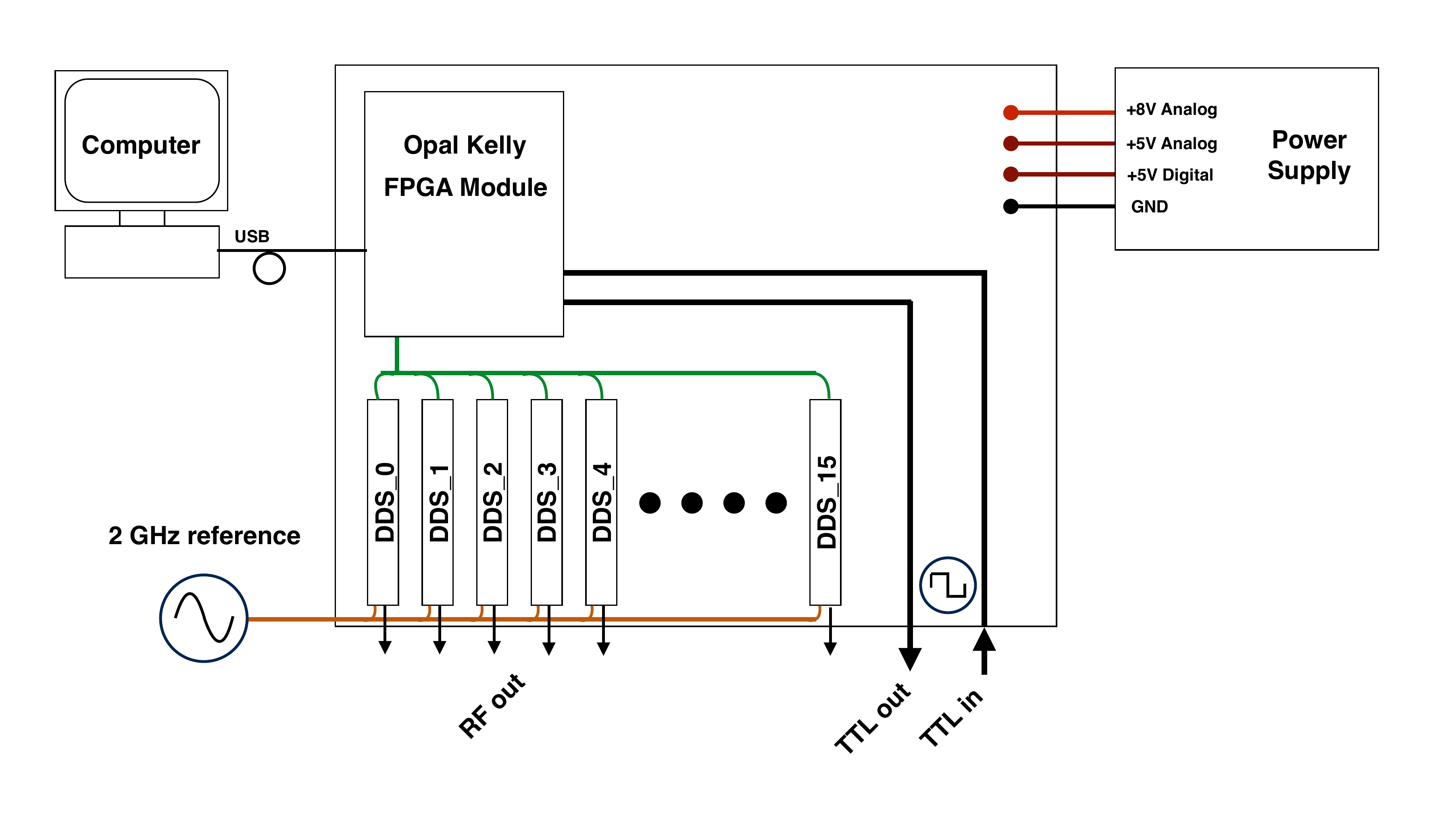}
\caption{Block diagram of a typical setup. The connection between the sequence unit and the computer is a single USB cable. Additional connections include a power supply and a reference clock at 2 GHz for each of the DDS channel. If more precise timing is required, we can supply a reference clock to the main FPGA using one of the TTL-input ports.}
\label{fig:block_diagram}
\end{figure*}

The block diagram of a typical setup of a complete system is shown in Fig.\ \ref{fig:block_diagram}. The core of the system is an FPGA module XEM6010 from Opal Kelly (OK FPGA), which connects to a user-interface computer via a single USB cable for data-transfer \cite{OK}. The OK FPGA unit controls the timing of the pulse sequence using its on-board oscillator \cite{timing_stability} (or an externally referenced clock via one of the TTL input channel) and outputs the digital TTL signals to control devices such as mechanical shutters or relays. For counting electrical pulses from a PMT, the OK unit receives a TTL signal from one of its input. 

To generate RF signals, the OK unit receives data (containing frequency, amplitude and phase for each DDS channel) from a computer and distribute the data to all 16 DDS boards. To increase flexibility, each DDS board also has a Cyclone IV (Altera) FPGA to store RF signal data and settings. Each FPGA on the DDS board then programs AD9915 (Analog Devices) DDS chip with desired frequency, phase and amplitude. Each DDS board requires a 2 GHz reference signal to operate properly. 

\begin{figure*}
\includegraphics[width=0.85\textwidth]{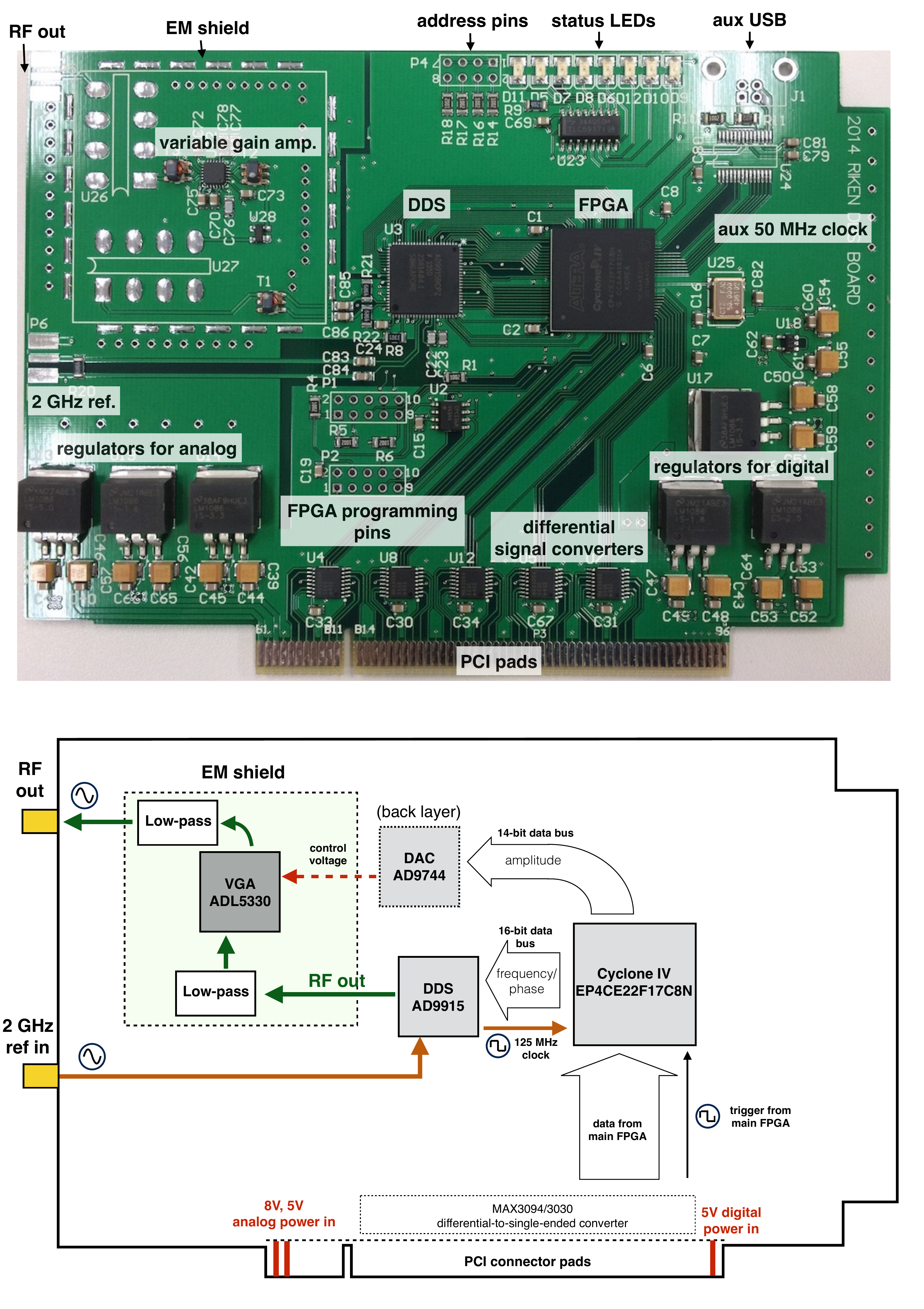}
\caption{\textbf{Top:} PCB of the DDS board with main components labeled. Some through-hole components and RF filters are not installed. \textbf{Bottom:} Block diagram of the DDS board. The Cyclone IV FPGA stores data received from the main OK FPGA and then sends the RF frequency and amplitude data directly to the DDS chip. To control the amplitude of the RF signal, the FPGA sends data to a DAC which then generates a control voltage to the VGA. Each DDS board requires a reference 2 GHz clock to operate.}
\label{fig:DDS_PCB_duo}
\end{figure*}

We use PCI connectors to attach each DDS board to the main PCB (see Fig.\ \ref{fig:DDS_PCB_duo}) for ease of installing and removing each individual board. The PCI bus contains a 16-bit bus and a few auxiliary signal lines for data-transfer between each DDS board and the main OK FPGA. Additionally, each DDS boards receives power directly from the PCI bus to simplify electrical connections. We can see in Fig.\ \ref{fig:DDS_PCB_duo} that the only connections are the 2 GHz input reference clock and an output RF signal for each DDS board.

\subsection{Digital pulses generator: the main Opal Kelly FPGA}

The main functionality of the main OK FPGA is to generate multi-channel (at least 32 channels) digital TTL pulses with timing programmable by the user. We define the data structure of the pulse sequence by the initial states of each TTL channel and the time in which each channel changes its state. In this way, memory used to store a pulse-sequence is determined by the complexity of the pulse sequence and not by the length of the pulse sequence (see Ref.\ \cite{Pruttivarasin2014a}). All the pulse sequence data is transferred into the memory of the OK FPGA before the starting of the pulse sequence. During the executing of the pulse sequence, the internal counter of the OK FPGA steps through the data stored in the memory. Then the TTL outputs change their states accordingly. Hence, there is no data transfer between the FPGA and the computer when the pulse sequence is running, eliminating any potential time delay during the pulse sequence. (The timing of the pulse sequence is determined faithfully by the on-board (or externally referenced) clock.) For our current design, the TTL pulses generator has a timing resolution of 40 ns. The switching time of the TTL signal is approximately 6 ns.

A separated counter in the OK FPGA is dedicated to counting electrical pulses from a PMT, which is widely used to detect fluorescence from trapped atoms or ions. The OK FPGA is able to time-tag the arrival of the PMT signal (with a timing resolution of 10 ns) relative to the start of the pulse sequence. The time-tagged data is stored temporarily in the internal memory of the FPGA which can be read out collectively later to reduce overheads in data-transfer. This is beneficial in running an experiment that requires long measurement time to build up statistics. This feature is demonstrated in Ref.\ \cite{Pruttivarasin2014b} where fluorescence from trapped ions is collected during many experimental runs before the all the time-tag data is read by the computer at the end of the pulse sequence.

Another functionality of the OK FPGA is to distribute RF signal data to all the DDS boards (described in the next section) via a 16-bit differential bus. Single-ended signals from the OK FPGA are converted to differential signals using MAX3030 and MAX3094 (Maxim) data converters to reject common-mode noise induced along the signal path. 

\subsection{RF generator: DDS board}

Each DDS board consists of a Cyclone IV FPGA and a DDS chip (AD9915 from Analog Devices). The block diagram is shown in Fig.\ \ref{fig:DDS_PCB_duo}. Before the starting of a pulse sequence, the FPGA receives data from the main OK FPGA and store it in a built-in memory in the FPGA. Each memory address in the FPGA is a 128-bit wide word used to store data for frequency, amplitude, phase, frequency ramping rate and amplitude ramping rate (see Table \ref{table:bit}). During the operation of the pulse sequence, the FPGA on the DDS board waits for a digital TTL pulse from the main OK FPGA to step to the next memory address. The FPGA then programs the DDS chip with the DDS data from this memory block.

Programming of frequency and phase is performed directly to the DDS chip via a 16-bit bus between the FPGA and the DDS chip (see Table \ref{table:bit}). For amplitude tuning, we control the variable-gain amplifier (VGA) ADL5330 (Analog Devices) which provides 60 dB dynamic range of amplitude tuning. The control voltage (0 to 1.2 V) for the VGA is generated from a high-speed AD9744 (Analog Devices) 14-bit digital-to-analog converter chip. This independent control over the frequency and amplitude allows us to perform frequency and amplitude ramping separately, which is impossible with the built-in digital ramp functionality of the DDS chip. We incorporate directly in the FPGA two independent counters with programmable counting rates. This allows us to generate more complex ramping patterns for other applications.

The layout of the printed circuit-board (PCB) of the DDS board is shown in Fig.\ \ref{fig:DDS_PCB_duo}. The PCB is a 4-layer board design with two inner planes used for ground and power. The analog part (left side) and the digital part (right side) have separated sets of voltage regulators to reduce electrical interference. There is also an auxiliary USB port for the purpose of using the DDS board as a stand-alone RF source.

\subsection{Control software}

Since the communication between the PC and the pulse-sequencer unit is done through the main OK FPGA module via a single USB cable, all data transfer is handled by the FrontPanel API provided by Opal Kelly. For us, we use the provided API in Python programming language. However, our experimental control software framework (LabRAD\cite{labrad}) allows an interface between various programming language, including LabVIEW. The details of the experimental control software is beyond the scope of this paper and we refer interested readers to Ref. \cite{Ramm2014} for full descriptions of the software. However, we would like to point out that to program a new pulse sequence to the OK FPGA, we do not have to recompile the hardware description code for the FPGA. The pulse sequence is written to the OK FPGA directly in Python data structure. 

An application software for controlling the pulse sequencer unit can be downloaded at a Git repository given in Ref. \cite{PulserGIT} which also includes design files for the PCB of the DDS board and VHDL source codes for all the FPGAs used in our setup.

\section{Performance}

\subsection{RF switching}

\begin{figure}
\includegraphics[width=0.45\textwidth]{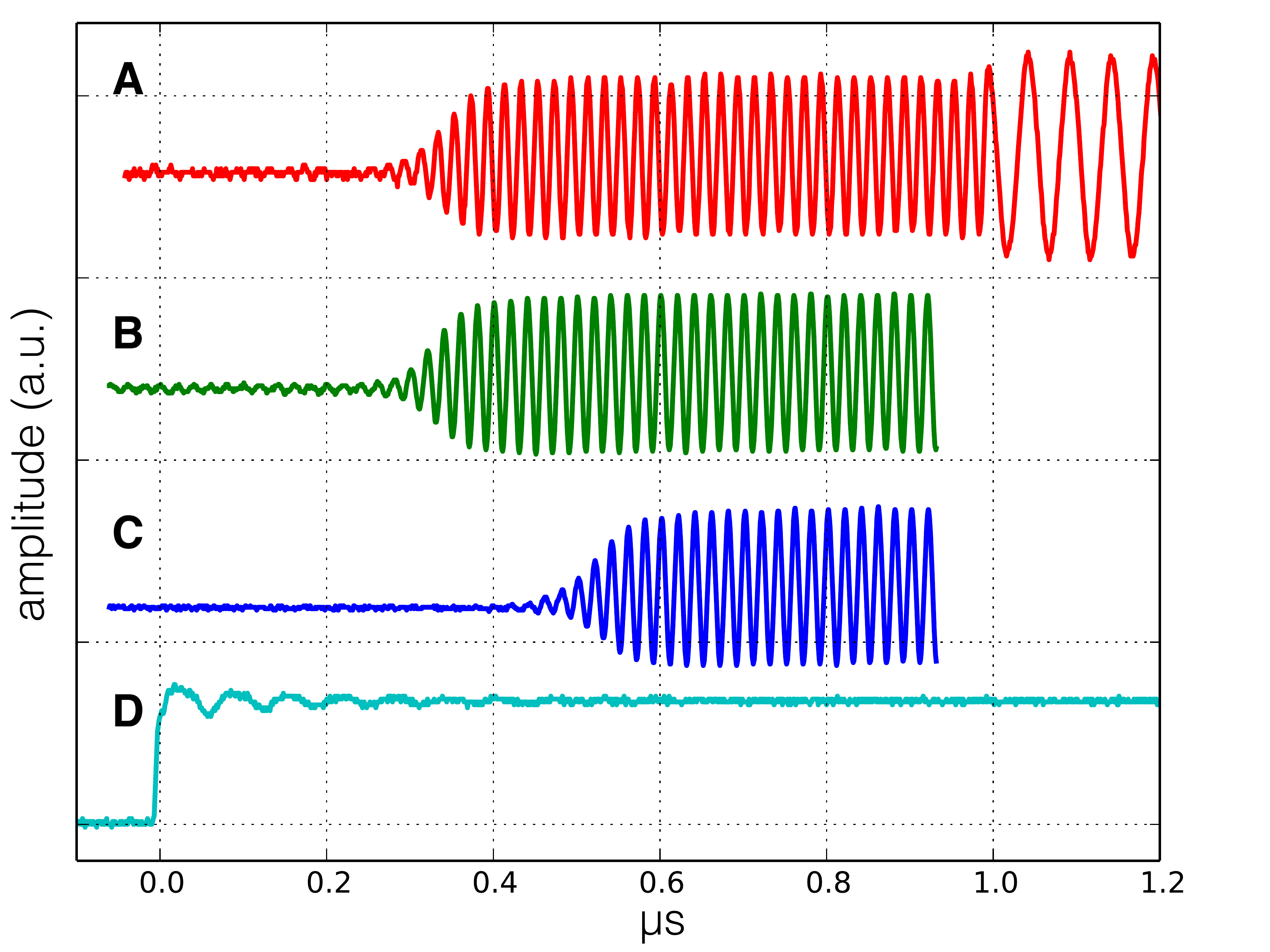}
\caption{Time required to switch the frequency and amplitude of the DDS channel. At t = 0.0 $\mu$s, a triggering pulse is sent to the DDS board from the main OK FPGA to update the DDS configuration. \textbf{Trace A} shows a case when both the amplitude and the frequency are changed at the same time. (Slight change in the RF amplitude after frequency switching is due to the response of the measuring oscilloscope.) \textbf{Trace B and C} show the case when the DDS switches the amplitude from completely off (C) compared to from a low but non-zero amplitude (B). \textbf{Trace D} is one of the TTL output for timing reference. }
\label{fig:trace_data}
\end{figure}

An ability to change the amplitude and/or the frequency of the RF signal (that drives the AOMs) rapidly is crucial in atomic physics experiments, especially in the case where the laser pulse duration is in the time scale of a few micro-seconds. Fig.\ \ref{fig:trace_data} shows the measured signal from one of the DDS channels where we set the main OK FPGA to trigger the DDS board at $t = 0.0~ \mu \text{s}$ (shown in Trace \ref{fig:trace_data}D) to switch both the amplitude (from low to high) and the frequency (Trace \ref{fig:trace_data}A and B). 

To understand the amplitude/frequency switching behaviour of the DDS channel as shown in Trace \ref{fig:trace_data}A, B and C, we describe here the protocol used for frequency/amplitude programming the DDS chip via the on-board FPGA (DDS FPGA). Once the DDS board receives a TTL trigger from the OK FPGA, the DDS FPGA advances its memory address to the next one and checks if there is a change in the amplitude and/or the frequency \cite{timing_note}. If there is only a change in the amplitude, the DDS FPGA sends data to the AD9744 DAC chip to update the control voltage to the VGA. (No data is sent to the DDS chip.) This task takes approximately 350 ns after triggering which is determined by the time to program the DAC chip and the response of the VGA. This is shown in Trace \ref{fig:trace_data}B.

However, if the amplitude changes to/from a completely off state, then in addition to the DAC chip, the DDS FPGA has to also program the DDS chip to turn off/on the RF output signal. This is shown in Trace \ref{fig:trace_data}C, where the RF amplitude is switched from a completely off state. This task takes approximately 200 ns longer to complete compared to Trace \ref{fig:trace_data}B because of the additional time it takes to program the DDS chip.

If there is a change in the frequency, then the DDS FPGA has to program the DDS chip with new frequency data. This task takes approximately 1.0 $\mu$s after triggering. In Trace \ref{fig:trace_data}A we change both the amplitude and the frequency. We can see that the amplitude switching is faster (and identical to Trace \ref{fig:trace_data}B) but the frequency switching takes longer to complete.

It is important to note that delay in programming is well-defined in terms of a number of clock cycles. The delay can be compensated in the control software. For example, if we want the frequency to switching exactly at $t$ = 0.0 $\mu$s, then we can trigger the DDS board at $t$ = -1.0 $\mu$s to obtain the desired timing.

\begin{figure}
\includegraphics[width=0.45\textwidth]{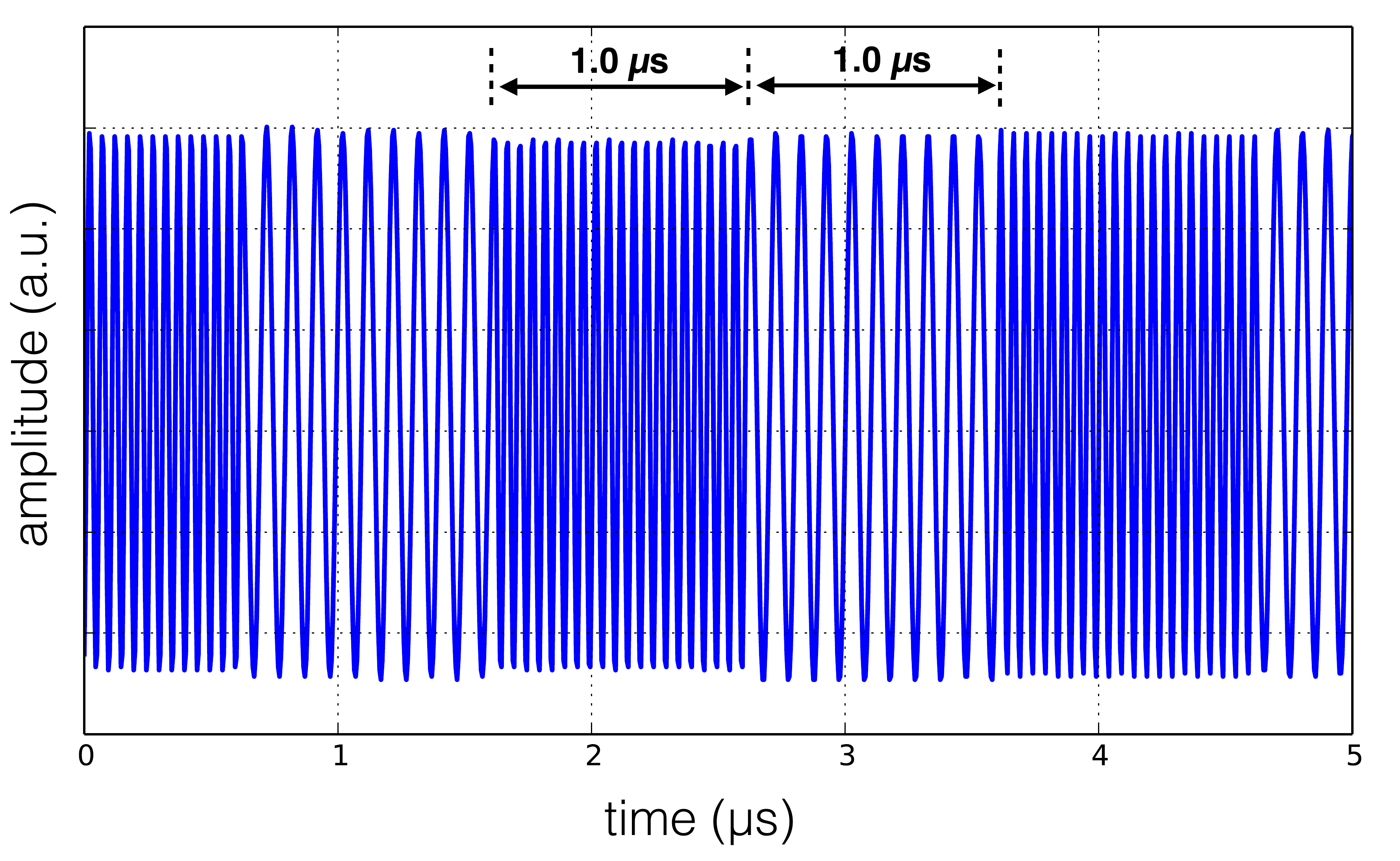}
\caption{Frequency switching of a DDS channel. Each DDS channel is capable of consecutively switching the frequency of the RF signal within 1.0 $\mu$s. This time is limited by the time the DDS FPGA takes to program frequency data to the DDS chip.}
\label{fig:freq_sw}
\end{figure}

We also demonstrate fast frequency modulation by alternating one of the DDS channel between two fixed frequency, as shown in Fig. \ref{fig:freq_sw}.

\subsection{RF phase control}

\begin{figure}
\includegraphics[width=0.45\textwidth]{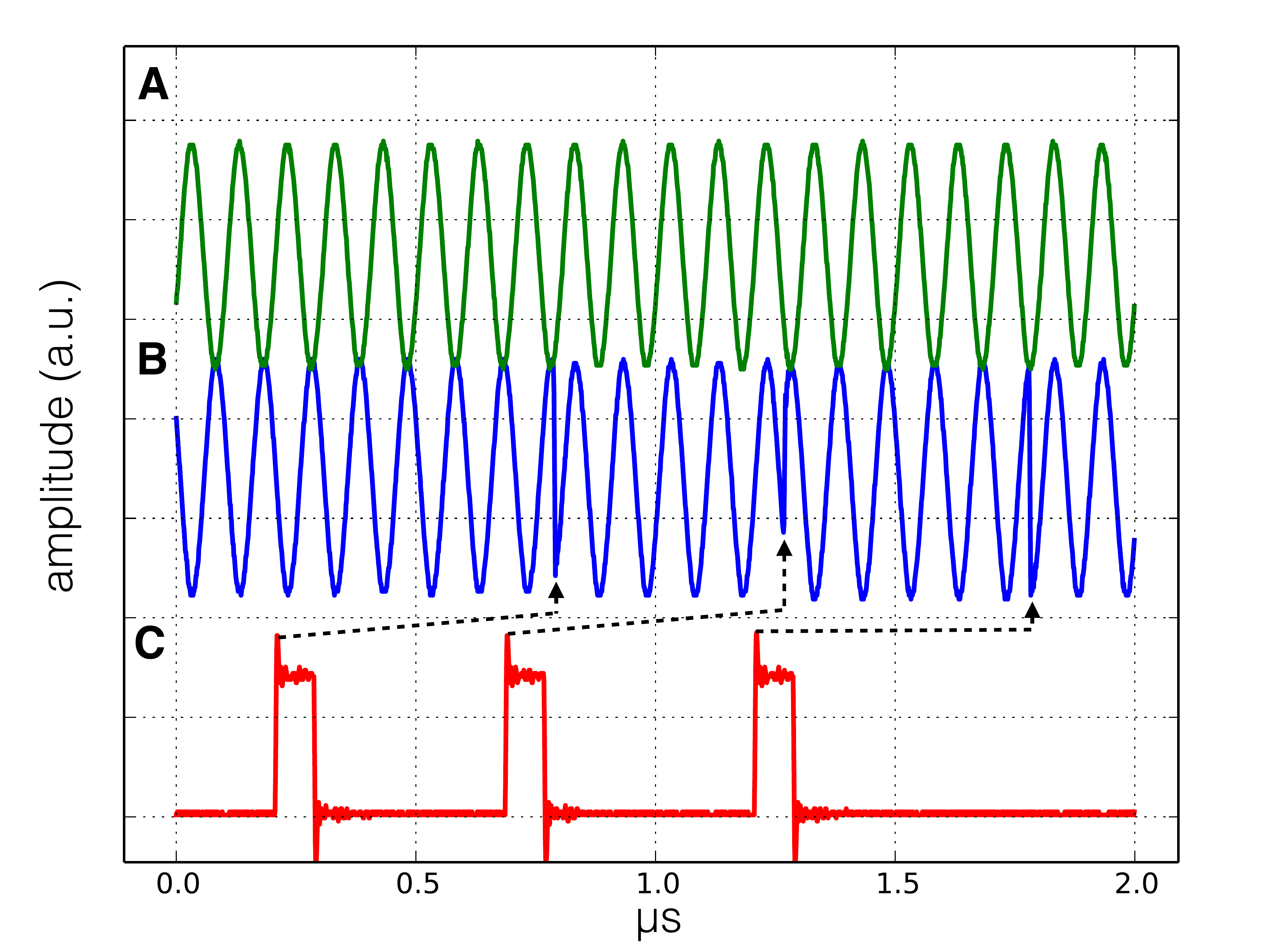}
\caption{Time required to switch the phase of the DDS channel. \textbf{Trace A} shows a reference RF signal. Each pulse in \textbf{Trace C} gives an update to the DDS channel (shown in \textbf{Trace B}) to flip the phase by 180\textdegree. The DDS takes $\sim500$ ns to update the phase of the RF signal.}
\label{fig:phase_data}
\end{figure}

In cold atom precision spectroscopy experiments, we often implement a Ramsey-type interferometric scheme. In this case, the phase of the laser light is directly controlled by the phase of the RF signal driving the AOM. The AD9915 DDS chip is capable of arbitrarily changing the phase of the RF signal by changing the phase offset in the internal phase accumulator. In Fig.\ \ref{fig:phase_data},  the main OK FPGA signals the DDS channel B to change the phase  by 180\textdegree~in three successive events given by digital pulses shown in Trace \ref{fig:phase_data}C. By comparing to a reference signal of DDS channel A, we can see that there is a delay of approximately 500 ns in phase switching. Since the delay is well-defined by a number of clock cycles the DDS FPGA takes to program the DDS chip, we can compensate this delay in a control software.

We successfully implemented the phase control capability in the work performed in Ref.\ \cite{Pruttivarasin2015}. In this work, a signal measured from trapped ions is used to feedback to the phase of the clock laser in a Ramsey-type interferometric scheme.

\subsection{RF amplitude and frequency ramping}

In some scenarios, fast amplitude and frequency switching of the RF signal driving AOMs are not desirable. For example, we might change the frequency of the laser light stabilized to a frequency comb by changing the reference frequency of the beat signal between the two. If the change in the reference frequency is too sudden, the stabilization circuit might not be able to follow. In this case, a slow change in the RF frequency is more desirable.

\begin{figure}
\includegraphics[width=0.45\textwidth]{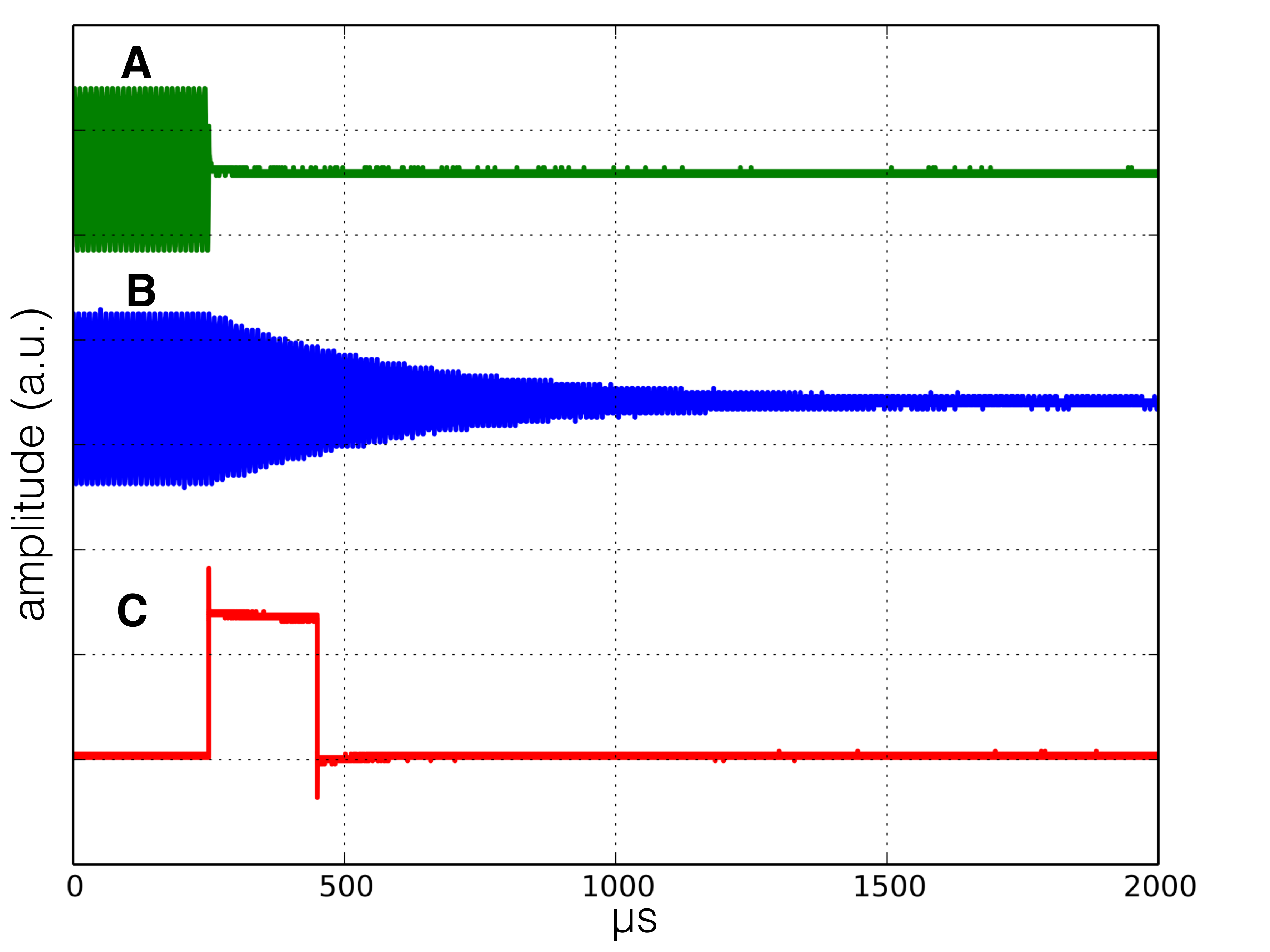}
\caption{Amplitude ramping. In \textbf{Trace A}, we simply switch off the RF signal while in \textbf{Trace B} we apply an amplitude ramping. Since the ramp is linear in the dB scale, the amplitude has an exponential decay profile. In this test, the ramping rate is set to be 20 dB/ms. \textbf{Trace C} is used for timing reference.}
\label{fig:amp_ramp_data}
\end{figure}

For each DDS channel, we implement a ramping capability for both frequency and amplitude by means of additional counters in the FPGA in each DDS board. Fig.\ \ref{fig:amp_ramp_data} shows a case where one of the DDS channel is configure to ramp down the amplitude (Trace B) compared to a sudden switching (Trace A). In this case, the ramping rate is set to 20 dB/ms.

\begin{figure}
\includegraphics[width=0.45\textwidth]{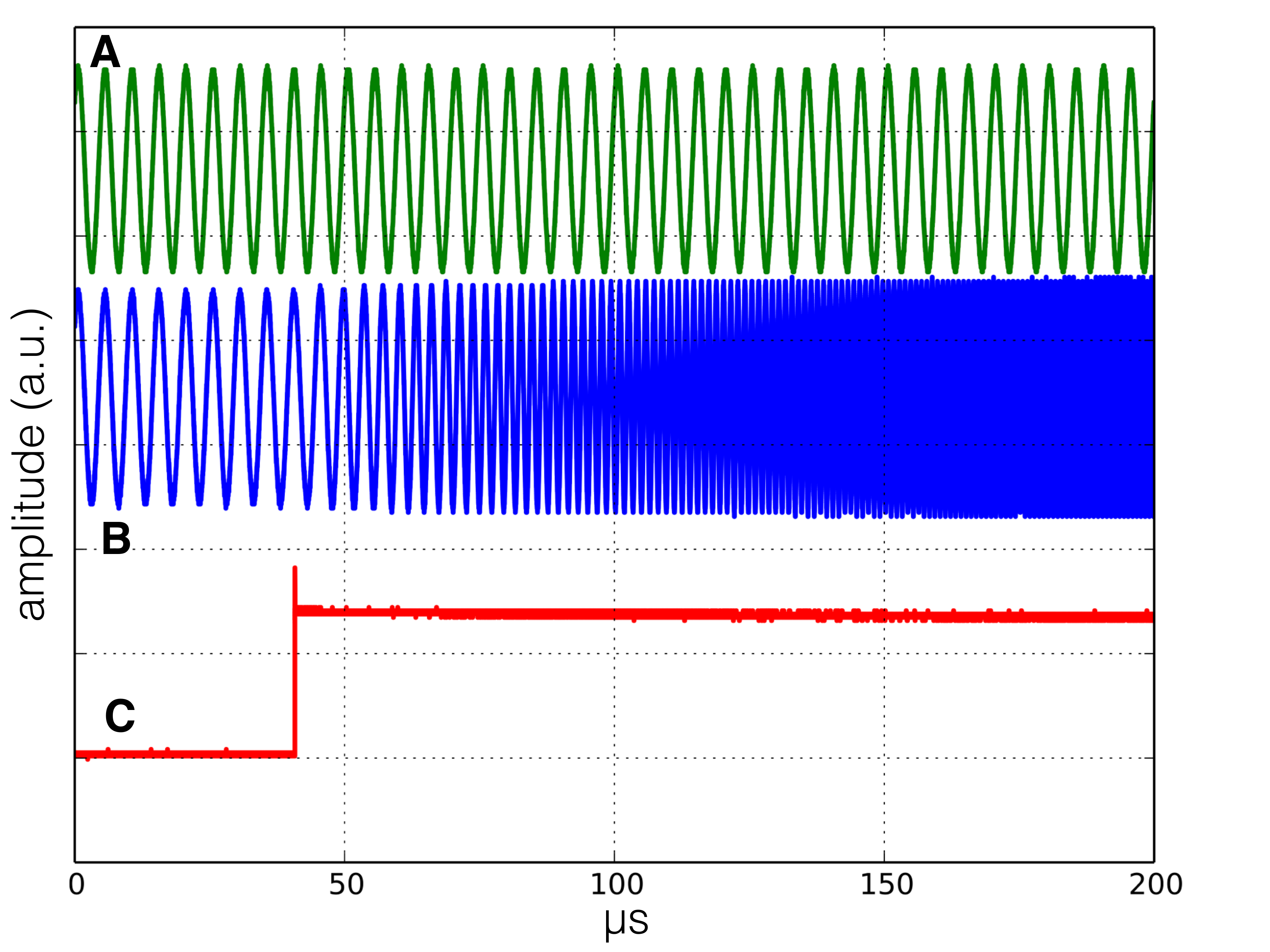}
\caption{Frequency ramping. In this test we apply a frequency ramping of 7 MHz/ms to one of the DDS channel (shown in \textbf{Trace B}). \textbf{Trace A and C} are used for frequency and timing references, respectively.}
\label{fig:freq_ramp_data}
\end{figure}

Fig.\ \ref{fig:freq_ramp_data} shows a capability of frequency ramping (Trace B) compared to a fixed RF frequency in Trace A. In this case the ramping rate is set to be 7 MHz/ms. We note that the DDS channel is capable of ramping both the amplitude and frequency simultaneously since they are two dedicated counters in the FPGA in each DDS board. We note that both frequency and amplitude ramping can be implemented at the same time.

\subsection{Sub-mHz frequency tuning}

Recent work on optical atomic clocks achieve the frequency precision in the mHz scale routinely\cite{Chou2010, Bloom2014, Ushijima2015}. It is then desirable to have RF sources that are capable of fine frequency tuning reaching the level below 1 mHz without sacrificing fast switching capability shown in the previous sections.

\begin{figure}
\includegraphics[width=0.45\textwidth]{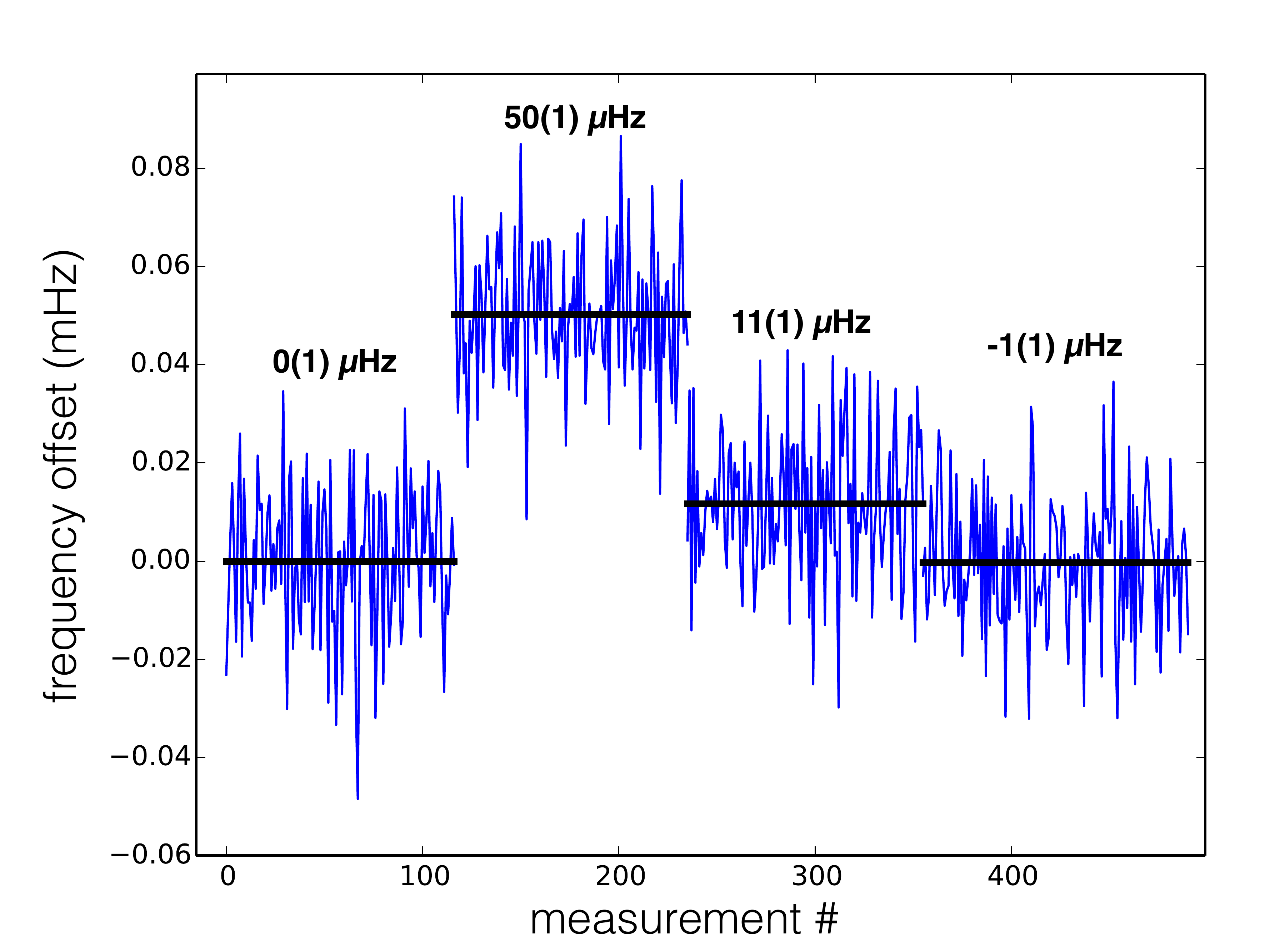}
\caption{Fine frequency tuning of the DDS. We set the frequency of one of the DDS channel to 0, 50, 10 and 0 $\mu$Hz offset from 15.22535454300 MHz and measure using a frequency counter Agilent 53230A.}
\label{fig:fine_freq_matome}
\end{figure}

In Fig.\ \ref{fig:fine_freq_matome}, we set one of the DDS channel to output frequencies of 0, 50, 10 and 0 $\mu$Hz offset from 15.225 354 543 00 MHz. To be able to resolve small frequency changes, we average the frequency readouts measured using a frequency counter (Agilent 53230A) for 5 minutes for each frequency setting. In this measurement, the 2 GHz reference clock for the DDS is referenced to the frequency counter. 

\subsection{Other tests}

\begin{figure}
\includegraphics[width=0.45\textwidth]{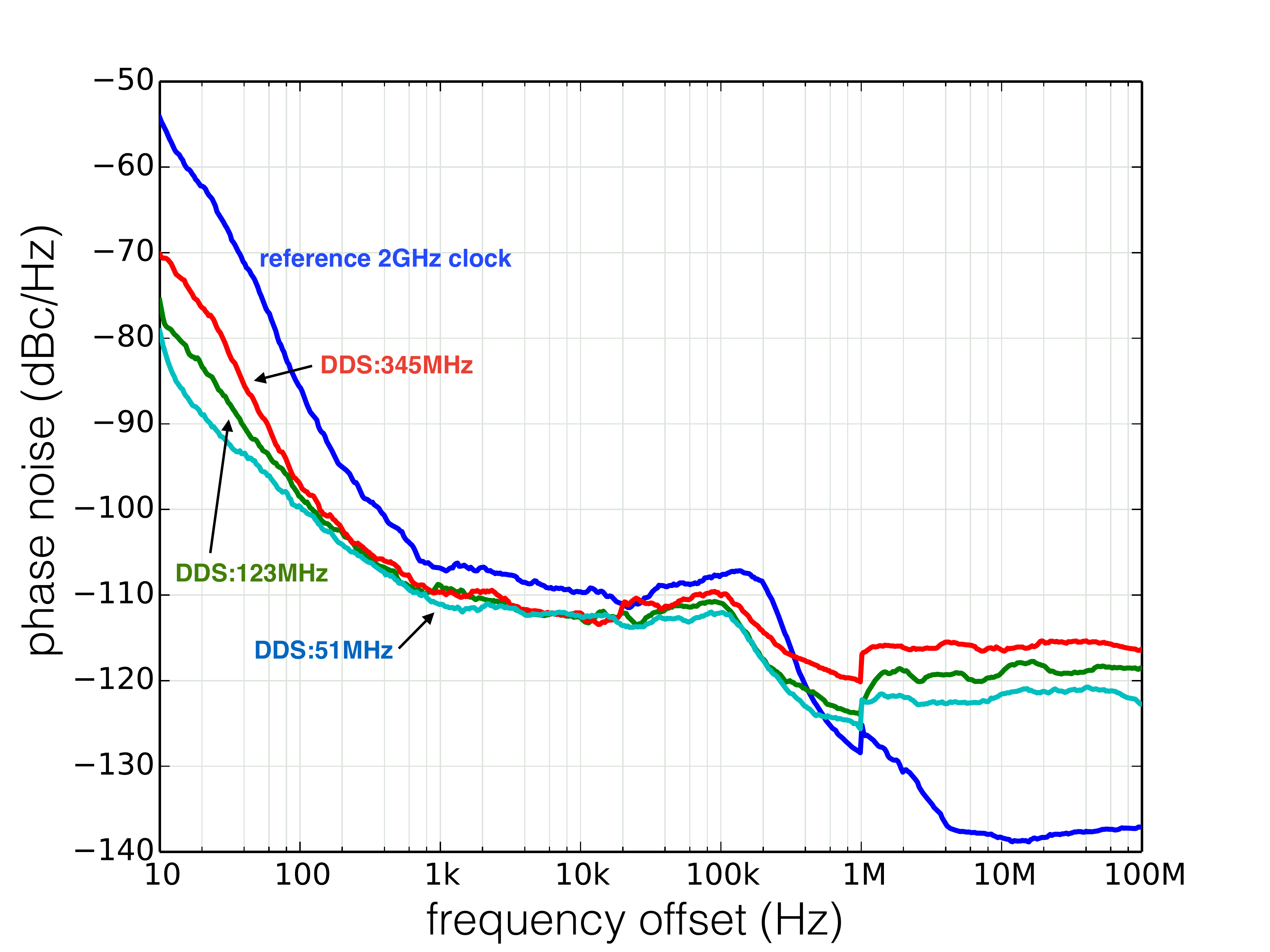}
\caption{Absolute phase noise of the DDS at various output frequency compared to the phase noise of the 2 GHz reference clock.}
\label{fig:phase_noise_data}
\end{figure}

\textbf{Phase noise:} We measured the phase noise of the DDS output and compared to the phase noise of the 2 GHz reference clock in Fig.\ \ref{fig:phase_noise_data}. We found that the noise relative to the reference clock is similar to the specification of the AD9915 chip\cite{AD9915}.

\textbf{Cross-talk:} We tested for cross-talk between two adjacent DDS channels by setting the RF power of one channel to maximum and looking for pick-up RF signal at the other channel (also set at maximum RF power). At the noise floor of $\sim120$ dBc of our spectrum analyzer, we did not see any RF pick-up in the adjacent DDS channel.

\begin{figure}
\includegraphics[width=0.45\textwidth]{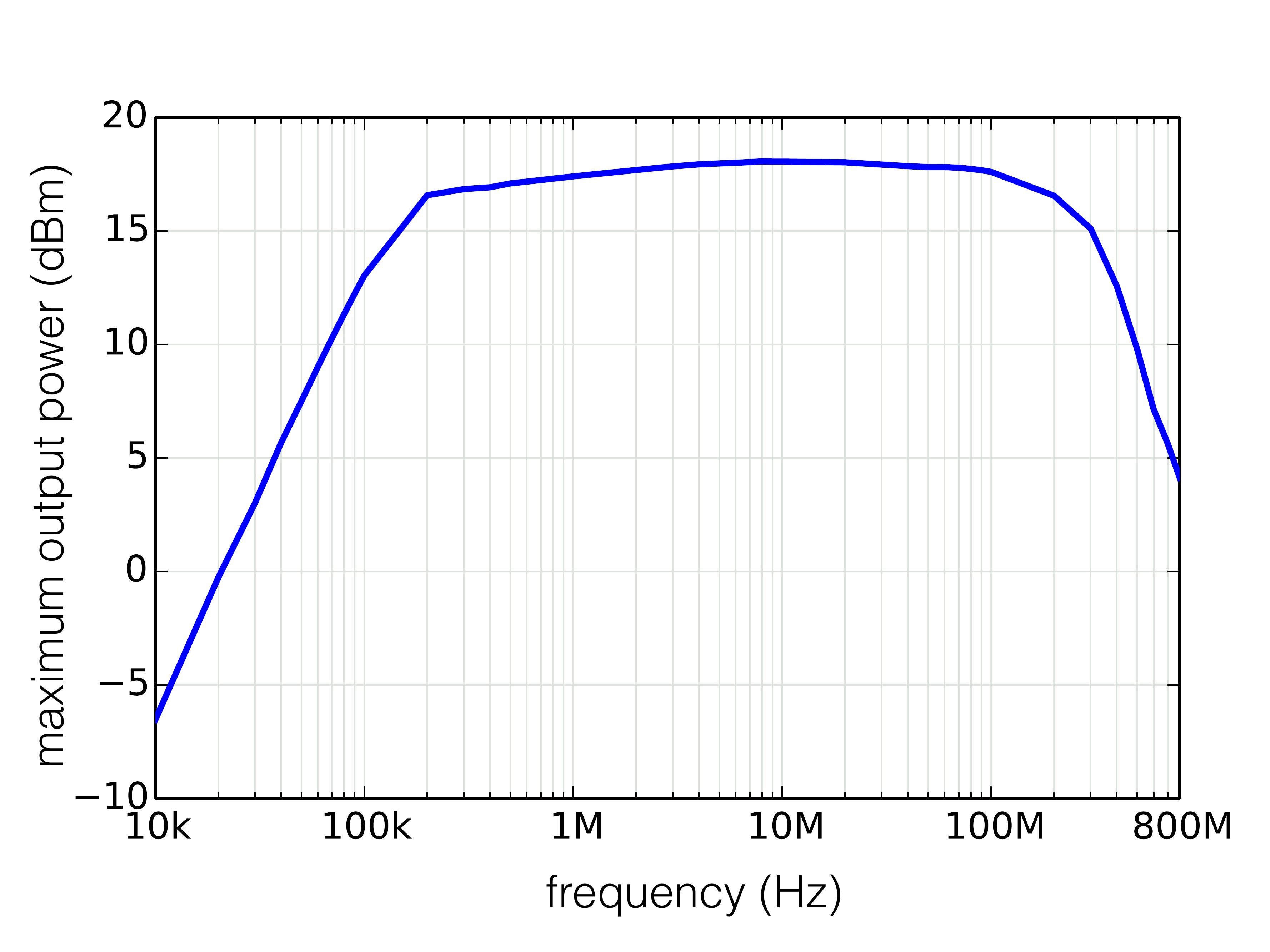}
\caption{Maximum output power for each DDS channel. The roll-off at low and high frequency is due to the bandwidth of the TC1-1-1T+ (Mini-Circuits) transformer.}
\label{fig:output_power}
\end{figure}

\textbf{Output power:} The output power as a function of the DDS frequency is shown in Fig.\ \ref{fig:output_power}. The roll-off at low and high frequency is due to a finite bandwidth of the TC1-1-1T+ (Mini-Circuits) transformer and ADL5330 variable gain amplifier on the DDS board. Data shown in Fig.\ \ref{fig:output_power} is taken without any on-board low pass filter.

\textbf{Power consumption and temperature:} for each DDS channel, the required current for the power supplies are approximately 400, 600 and 150 mA for 5V (digital), 5V (analog) and 8V (analog), respectively. Without active air-flow cooling and room temperature of 25\textdegree~C, the DDS chips heats up to approximate 45\textdegree~C during a normal operation.

\section{Summary}

We have presented a pulse-sequencer and RF generators unit suitable for experiments in atomic physics where the amplitude and frequency of laser lights are controlled by AOMs and mechanical shutters. Sub-mHz frequency tuning of the RF generators makes the system suitable for optical atomic clock, where mHz frequency resolution of the laser light frequency is routinely achieved. Additionally, the timing of the pulse-sequence, ranging from sub-micro-seconds to seconds, together with frequency and amplitude ramping functionality, adds robustness to the pulse sequencer to be applicable to a wide variety of experiments with trapped atoms and ions.

\section{Acknowledgements}

T. P. would like to thank H. H{\"a}ffner and M. Ramm for support and assistance during the development of the pulse sequencer system in Berkeley. This work is supported by RIKEN's Foreign Postdoctoral Researcher program.

\begin{table*}
\caption{Allocation of DDS data in each of the 128-bit memory block in the FPGA.}
\label{table:bit}
\begin{tabular}{lccc}
\hline\hline
Type of data & No. of bit  & Resolution   & Range    \\ \hline
Frequency & 64    & (2.0/$2^{64}$) GHz = 0.1 pHz & 0 to 800 MHz \\
Amplitude  & 14    & (60/$2^{14}$) dB = 0.004 dB & -60 to 0 dBm \\
Phase & 16    & (360/$2^{16}$)\textdegree = 0.0055\textdegree& 0 to 360\textdegree \\
Frequency ramp & 16    & (7.45/$2^{16}$) MHz/ms = 113 Hz/ms & up to 7.45 MHz/ms\\
Amplitude ramp & 16   & 0.0017 dB/ms & up to 22.9 dB/ms \\
\hline
\end{tabular}

\end{table*}

\appendix


\begin{thebibliography}{000}
\bibitem{timing_note} A triggering TTL pulse from the main OK FPGA is sampled by a clock on each DDS board. Hence, jitter in the triggering timing for each DDS board is less than one period of the clock on the DDS board. This is on the order of 10 ns for our system. This jitter can be reduced further by referencing the clock used by the OK FPGA to the same one as the DDS board.
\end{thebibliography}
\end{document}